\documentclass[journal,twocolumn]{IEEEtran}
\IEEEoverridecommandlockouts                              % This command is only
\newtheorem{theorem}{Theorem}

\newtheorem{lemma}[theorem]{Lemma}

\RequirePackage{amsmath, amssymb, graphicx, url,
xspace,subfigure,braket,epstopdf}
\usepackage{comment} 
\usepackage[makeroom]{cancel}

\usepackage{graphicx,subfigure}% for pdf, bitmapped graphics files
\usepackage{epsfig} % for postscript graphics files
\usepackage{mathptmx} % assumes new font selection scheme installed
\usepackage{times} % assumes new font selection scheme installed
\usepackage{amsmath} % assumes amsmath package installed
\usepackage{amssymb}  % assumes amsmath package installed
\usepackage{amsfonts}
\usepackage{euscript}
\usepackage{color}
\begin{document}

\title{\LARGE \bf Sliding-mode theory under feedback constraints\\ and the problem of epidemic control}
\author{Mauro Bisiacco and Gianluigi~Pillonetto
        \thanks{Mauro~Bisiacco (bisiacco@dei.unipd.it) and Gianluigi~Pillonetto (giapi@dei.unipd.it) are with Dipartimento di Ingegneria
        dell'Informazione, University of Padova, Padova, Italy. 
%This paper was
%not presented at any IFAC meeting. 
Corresponding author Gianluigi
Pillonetto Ph. +390498277607}}
%\author{Anna~Scampicchio
%        \thanks{A. Scampicchio (anna.scampicchio@gmail.com) is with Dipartimento di Ingegneria
%        dell'Informazione, University of Padova, Padova, Italy.}}

\maketitle \thispagestyle{empty} \pagestyle{empty}

%%%%%%%%%%%%%%%%%%%%%%%%%%%%%%%%%%%%%%%%%%%%%%%%%%%%%%%%%%%%%%%%%%%%%%%%%%%%%%%%
\begin{abstract}
One of the most important branches of nonlinear control theory 
is the so-called sliding-mode. Its aim is the design of a 
(nonlinear) feedback law that 
brings and maintains the state trajectory of a dynamic system on a given sliding surface.
Here, dynamics becomes completely independent of the model parameters and can be tuned accordingly to the desired target.
In this paper we study this problem when 
the feedback law
is subject to %some 
strong structural constraints. 
In particular, we assume that the control input may take values only over two bounded and
 disjoint sets. Such sets could be also non perfectly known a priori. 
An example is a control input allowed to  
switch only between two values. 
Under these peculiarities, we derive the necessary and sufficient conditions 
that guarantee sliding-mode control effectiveness for a class of time-varying 
continuous-time linear systems that includes all the stationary state-space linear models.
Our analysis covers several scientific fields. It
is only apparently confined to the linear setting and allows also to 
study an important set of nonlinear models.
We describe fundamental examples related to epidemiology 
where the control input %may be represented by 
is the level of contact rate among people
and the sliding surface permits to control the number of 
infected. % or of people in intensive care.
For popular epidemiological models we prove the global convergence of control schemes
based on the introduction of severe restrictions, like
lockdowns, to contain epidemic. 
This greatly generalizes previous results  
obtained in the literature by casting them within a general sliding-mode theory. \end{abstract}
%\acks{This research has been partially supported by the
%Progetto di Ateneo CPDA147754/14-New statistical learning approach
%for multi-agents adaptive estimation and coverage control.
%This paper was
%not presented at any IFAC meeting. Authors would like to thank 
%Alexander Goldenshluger for helpful discussions on the arguments developed in this paper. 
%Corresponding author Gianluigi
%Pillonetto Ph. +390498277607.}
%%%%%%%%%%%%%%%%%%%%%%%%%%%%%%%%%%%%%%%%%%%%%%%%%%%%%%%%%%%%%%%%%%%%%%%%%%%%%%%%
%%%%%%%%%%%%%%%%%%%%%%%%%%%%%%%%%%%%%%%%%%%%%%%%%%%%%%%%%%%%%%%%%%%%%%%%%%%%%%%%

\section{Introduction}

Dynamic systems play a prominent role 
in modern science. % and, %In fact, many physical phenomena
%surrounding us can be seen as objects that vary in time.
%They are subject to (desired or non desired) inputs and produce observable signals called outputs.
Within this broad concept, two key problems arise in many real-world applications.
The first one is inferring mathematical models able to suitably reproduce %the working of 
the system through experiments where input-output data are collected,
a task known as system identification in the engineering literature \cite{Ljung:99,PillonettoDCNL:14}. 
%and typically returns a nominal model and uncertainty bounds around it.
The second one is concerned with control \cite{Astrom2014}, a problem that 
typically requires the design of feedback laws that make
the system evolve according to the desired behaviour.
One way to obtain this goal is to resort to the so-called 
sliding-mode technique  \cite{Sliding1992,Sliding1998}.
It relies on the design of discontinuous
control inputs and it will be the focus of this paper. 
Such approach represents one
of the most important branches of nonlinear control theory 
\cite{Isidori1995}, with many applications in several contexts,
ranging from industry, robotics \cite{Sliding2012,SlidingBook} and biosciences 
where positive systems are often encountered \cite{Ren2018}.\\ 

When adopting sliding-mode controllers, 
the desired system behaviour is encoded in the choice of a sliding manifold
that thus defines the control objective.
A discontinuous control law is then designed 
in order to reach the manifold in finite-time
and to maintain the system state confined to it. 
Here, dynamics become completely independent of the model parameters
and a suitable equilibrium point can be made asymptotically stable.
Hence, the desired control target can be satisfied.
Another important feature of sliding-mode control is also 
its robustness against the system uncertainties that inevitably affect the nominal model
%and are typically 
returned by the system identification procedure.\\

The fundamental novelty present in this work is that
we study sliding-mode control assuming 
that the structure of the feedback law is subject to 
strong constraints. We consider 
a situation where  
the control input may assume values in two bounded and non overlapping
sets that can be also unknown. % (that can be also 
Hence, our analysis includes also the case
of an input that can switch only between two values.
Under these restrictions,
we obtain the necessary and sufficient conditions that %still 
guarantee the effectiveness of the control
for a class of time-varying linear systems in continuous-time that includes all the stationary state-space models \cite{Kailath79}.
It will be proved that such conditions involve system controllability %controllability 
and the product of the determinants of two matrices.
%The result is then specialized
%to a class of positive systems where the state variables can assume only nonnegative values.
%Finally, 
Even if apparently confined to the linear setting, we illustrate how the analysis includes also an important class of 
nonlinear models. In fact, one can exploit 
 the time-varying component of the system
to capture nonlinear dynamics.
This makes the characterization of the convergence and stability properties of the proposed family of controllers relevant in 
many different contexts like epidemiology, as described in the next section.\\

The paper is organized as follows. 
Section \ref{MotEx} illustrates a motivating example regarding epidemic control. 
Section \ref{Res} first reports our theoretical findings on
sliding-mode under feedback constraints. Then, the new results are used to gain new insights  
on the problem of epidemic control, also revising the motivating example introduced in the previous section.
Conclusions end the paper while proofs of the mathematical results are gathered in Appendix.
 
%Finally, it is shown that Despite it could seem that (non-linear) epidemiological models are outside this setting, the time-variance of the parameter will allow us to easily include them as particular cases\\
%i sistemi positivi
\section{A motivating example: control of an epidemic}\label{MotEx}
%The theoretical analysis on the convergence and stability properties of the proposed family of controllers 
%is relevant in several scientific fields including e.g. biomedicine and epidemiology.
The motivation %that we will return to later and che riprenderemo anche piu' avanti
that prompted us to undertake this study is related to 
the COVID-19 pandemic that first appeared in Wuhan, China
\cite{Fei2020,Wu2020,Guan2020} and
then spread all over the world
 \cite{Velavan2020,Wittkowski2020}.
 Under the impact of COVID-19  emergency, 
modeling and control of epidemic models
has been recently subject of new and intensive research \cite{Gatto2020,Giordano2020,CCGondim2020,CCTsay2020,CCKohler2020,CCBerger2020,pillonetto2020tracking}.
In particular, we are interested in the theoretical study of those
control strategies adopted %by many countries 
to contain the epidemic and 
%(before the development and administration of vaccines)
based on social distancing measures, including also 
strong restrictions in the form of lockdown \cite{Crisanti2020,VirusVar2021}.
To describe the control problem,
just for the sake of simplicity, we start considering the SEIR model \cite{Bootsma2007,Capasso1978,Korobeinikov2005,Liu1987}.
It is an example of compartmental model, where the population is assumed to be
well-mixed and divided into categories. SEIR represents 
one the of most popular generalizations of the SIR model \cite{Kermack1927,Bertozzii2020} and
includes also (exposed) people who are host for infection but cannot yet transmit the disease.
In particular, four classes $S(t),I(t),E(t)$ and $R(t)$ evolve as function of time $t$ and 
contain, respectively, susceptible, infected, exposed and removed 
people. They are normalized, hence their sum is equal to one for any temporal instant $t$, and obey 
the following set of differential equations % then define the \emph{SEIR model} %which we call SIIR %modified SEIR   
 \begin{subequations}\label{SEIReq}
\begin{align}
\dot{S}(t)&= - \beta(t)S(t)I(t) \\% \ a>0; \\
\dot{E}(t)&= \beta(t)S(t) I(t) - \epsilon E(t) \\ %a(t)=\beta(t)
\dot{I}(t)&= \epsilon E(t)- \delta I(t) \\ %, \ b>0; \\
\dot{R}(t)&= \delta I(t)%\\
%y(t)&=&\frac{1}{H}I(t). 
\end{align}
\end{subequations}
% where $c$ is the rate with which infected people $I(t)$ recover
%and 
where %, scrolling through the equations from the bottom, 
the scalar $\delta$ is the rate with which infected people heal or die while
$\epsilon$ is the rate with which exposed become infected.
Finally, the time-varying variable $\beta(t)$ is the infection rate that 
accounts for the transmissibility/contagiousness of pathogens agents 
by describing the interaction between susceptible and infected.
Its value depends not only on the biological characteristics of the virus but also 
on all those factors which influence the human behaviour including 
social organization.\\
During an epidemic,  $\beta(t)$ can be seen as a
control input, manipulable (to some extent) 
through preventive and interventional measures introduced on the basis  
of the number of infected people \cite{WAtoday2021}.
It is not however possible to implement a control law where  
$\beta(t)$ is a continuous function of $I(t)$.
In practice, only when the number of infected 
enters a certain range new restrictions are set or removed.
%hence alternating different periods of freedom and lockdown/reduced freedom.
One can instead think that $\beta(t)$ may assume values over two 
bounded and non overlapping sets. They are not even known a priori:
restrictions of different shades of intensity are typically introduced 
and their  influence on the contact rate is never perfectly predictable.  
What is know is only that the feedback law changes the value of $\beta(t)$ 
by alternating periods of freedom and lockdown.
Interestingly, under the stated constraints on the contact rate,
global convergence of sliding-mode controllers of $I(t)$ applied to the SEIR, 
i.e. the ability to control epidemic starting from any initial condition,
has never been demonstrated. 
The same also holds when other models
of COVID-19 dynamics are adopted.
An example is the SAIR where the class of exposed is replaced with 
that of asymptomatic people \cite{Sadeghi2021} 
who are known to play a relevant role in transmitting COVID-19 
\cite{Wang2020,Crisanti2020}. A more sophisticated model that will be 
also reported later on is the 
SEAIR that includes both of these classes.
The analysis developed in this paper will fill this gap by providing the necessary and sufficient conditions for epidemic control,
generalizing previous results  
obtained in the literature like \cite{Ibeas2013,Xiaoqi21,Menani2017,Nunez2021,AnnalsSliding2021}. 
These latter will be in fact cast within a more general framework
where the control of models like SEIR, SAIR or SEAIR becomes just a special case 
of a general sliding-mode theory that always guarantees asymptotic stability. 

\begin{figure*}[h]
	\begin{center}
		\begin{tabular}{c}
			\hspace{-.2in}
			{ \includegraphics[scale=0.45]{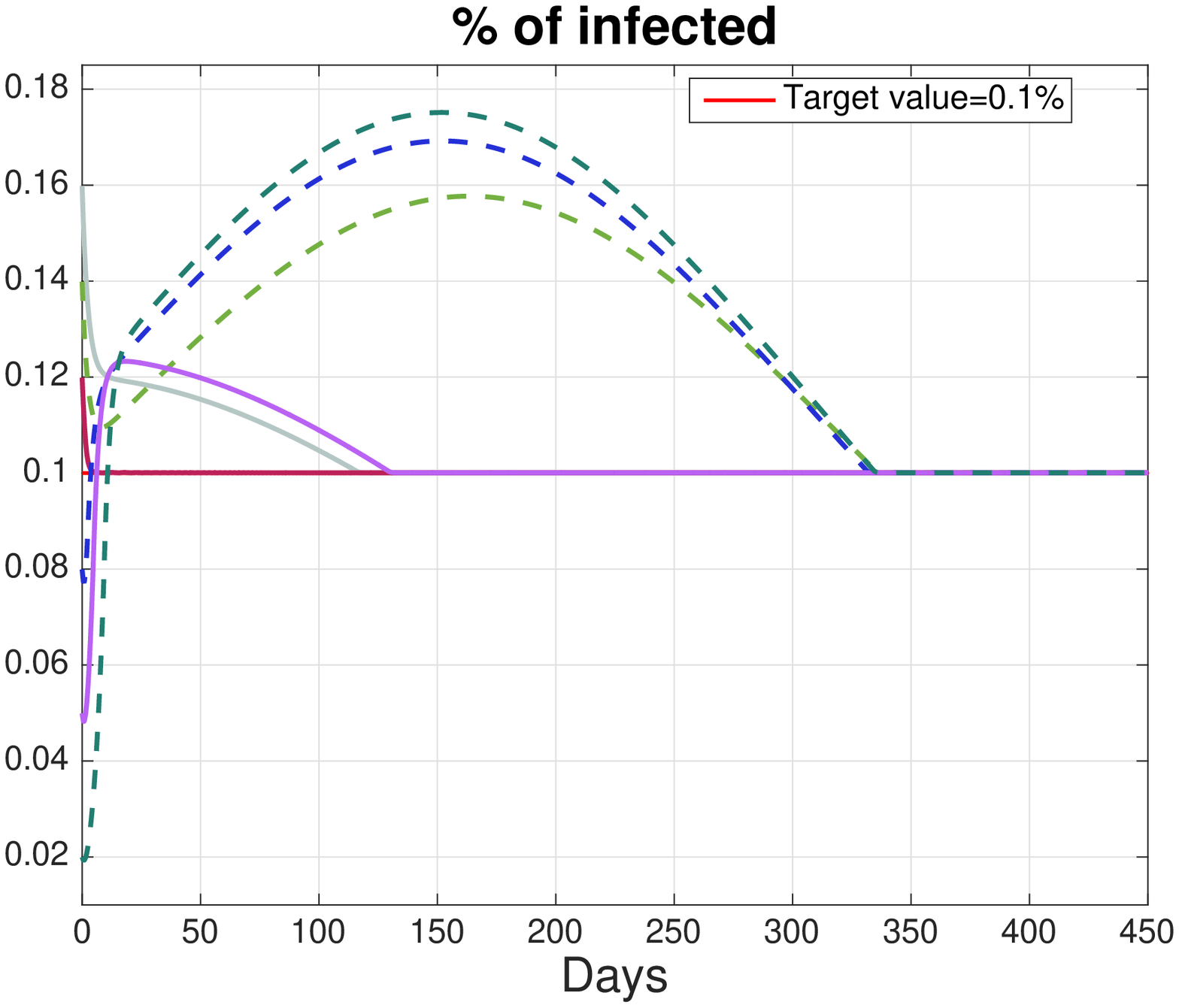}} \ { \includegraphics[scale=0.45]{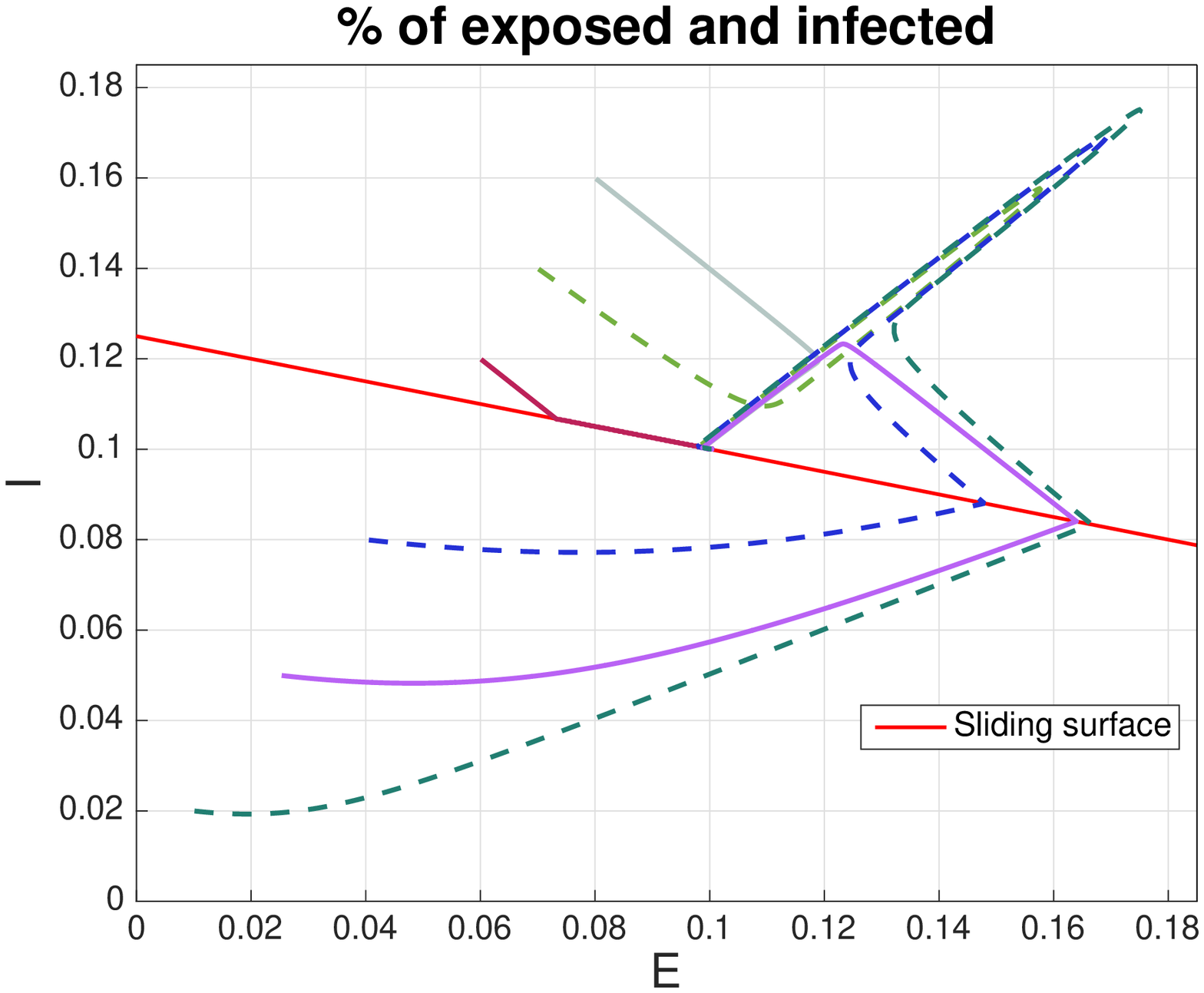}}
		\end{tabular}
		\caption{{\bf Motivating example: epidemic control} The theory developed in this paper proves the convergence of sliding-mode controllers applied to a wide class of models. An example is the control of epidemic evolution as described e.g. by means of the SEIR model reported in \eqref{SEIReq}. The target is to control the number of infected maintaining it to the desired value $I_0$. This can be done by influencing the time-course of the contact rate $\beta(t)$ through interventional measures that limit social interaction. 
		As discussed later on, in this case a convenient  sliding-mode surface is defined by $\epsilon E-\delta I+\lambda(I-I_0)=0$ where $\lambda>0$ 
		and the (discontinuous) feedback law sets a lockdown if $\lambda(I-I_0)+\dot{I}>0$. 
		For simplicity, we can now think that two values for the contact rate $\beta(t)$ are alternated, equal to $\beta_F$
		when no restrictions are set and to $\beta_L$ during the lockdown.
		The theory developed in this paper provides the necessary and sufficient conditions 
		ensuring that, for any system initial condition,
		the equilibrium point is reached and maintained (until the epidemic dies out).
                 Some time-courses of the number of infected generated by the closed-loop system
                 are shown in the left panel. 
                 They all converge to the target value $I_0$ that is set to $0.1\%$ of the population (horizontal line).
                 The right panel also displays the corresponding trajectories over the exposed-infected plane 
                 and the sliding region (red straight line).          
                 Adopted parameter values are $\beta_F=0.8$, $\beta_L=0.2$ (associated trajectories are plotted using solid lines) or $0.21$ (dashed lines), $\delta=\epsilon=0.2$, $\lambda=1$.
                 Attractiveness of the sliding region visible in figure is the manifestation of the general sliding-mode
                 convergence theory under feedback constraints developed in the paper. Examples treated in what follows will regard also other epidemiological models, like SAIR and SEAIR, that incorporate the class of asymptomatic people who are known to play a major role in transmitting COVID-19.} \label{Fig1}
	\end{center}
\end{figure*}

\section{Results} \label{Res}

\subsection{Sliding-mode convergence theorem under feedback contraints}

The following theorem represents the main result of this paper.
It considers a state-space $n$-dimensional linear model whose input is defined by a state feedback %from the state
subject to a time-varying gain $\gamma(t)$. Such gain is uniformly bounded in time,
a constraint defined by the union of two disjoint sets $I_1$ and $I_2$. 
%We would like 
The control target is to make 
systems dynamics follow the Hurwitz polynomial $\Delta(s)$ of degree $n-1$.
The necessary and sufficient conditions for a sliding-mode controller to satisfy such
requirement are then obtained (the proof is reported in Appendix).\\

\begin{theorem}\label{Main} Consider the following single-input linear system with $\dim(x)=n$
$$
\dot{x}=Fx+gu, \ u=Hx\gamma(t),
$$
where $\gamma(t)$ is a bounded time-varying gain. 
We are also given a  pair of disjoint intervals 
$I_1=[\gamma_1^{(1)}, \gamma_1^{(2)}], I_2=[\gamma_2^{(1)}, \gamma_2^{(2)}]$, with $\gamma_1^{(2)}<\gamma_2^{(1)}$,
such that $\gamma(t) \in I_1 \cup I_2$ for any $t \ge 0$, and an
arbirary $(n-1)-$degree Hurwitz polynomial $\Delta(s)=s^{n-1}+\dots+\Delta_1s+\Delta_0$.
%\medskip

\begin{center} {\bf Then} \end{center}
\medskip

\noindent there exist a unique matrix $K$, a unique $\gamma_0 \in [\gamma_1^{(2)}, \gamma_2^{(1)}]$ and a point $x_{eq} \ne 0$ (univocally defined except for a multiplicative constant) such that
\begin{itemize}%[leftmargin=1cm]
\item $[F+gH\gamma_0]x_{eq}=0$;
\item  the following control law
$$
\begin{array}{lcl}
\gamma(t) &=& \ \mbox{any value} \ \in I_1 \ \ \mbox{if} \ \ (Hx_{eq})[K(x-x_{eq})]<0, \cr
\gamma(t) &=& \ \mbox{any value} \ \in I_2 \ \ \mbox{if} \ \ (Hx_{eq})[K(x-x_{eq})]>0
\end{array}
$$
leads to the sliding surface $Kx=Kx_{eq}$, endowed with dynamics associated to the characteristic polynomial $\Delta(s)$;
\item the sliding surface is (at least) locally attractive in a suitable open neighborhood ${\mathcal I}$ of $x_{eq}$. Hence, 

\begin{itemize}%[leftmargin=1cm]
\item ${\mathcal I}$ is invariant, i.e. $x(0)\in{\mathcal I}$ implies $x(t)\in{\mathcal I}$, for any $t\ge 0$;
\item $x(0)\in{\mathcal I}$ implies that $x(t)$ reaches the sliding surface in finite time. From that time onwards $x(t)$ does not escape from it and then $x(t)$ tends to $x_{eq}$ for $t$ tending to $+\infty$,
\end{itemize}
\end{itemize}

\medskip
%$\Big \Updownarrow$ \\
\begin{center} {\bf if and only if} \end{center}
\medskip

\noindent the following conditions hold true
\begin{itemize}%[leftmargin=1cm]
\item $(F,g)$ is a controllable pair; %controllable
\item $\det[F+gH\gamma_1^{(2)}]\det[F+gH\gamma_2^{(1)}]<0$.
\end{itemize}

\end{theorem}
%\hfill$\heartsuit$
%\medskip

%\noindent {\bf Remark.} \ Obviously, the theorem statement ensures the asymptotic stability of $x_{eq}$, in a local sense, too. However, nothing can be inferred in general about the (possible) {\bf global asymptotic stability} of $x_{eq}$ itself. \hfill$\heartsuit$.

\subsection{Applications to epidemic control}

As already anticipated, despite the apparent linear assumptions, the applicability of the theorem 
extends well beyond this setting. To this regard, the time-varying gain $\gamma(t)$
plays a crucial role since it can be used to capture also hidden nonlinear parts of a system.
This fact will become clear through the following illustrations regarding epidemic control. 
We start treating the SEIR  model that was the basis of our motivating example.

\subsubsection{SEIR}

Consider now  \eqref{SEIReq}, define 
$$
\gamma(t):=\beta(t)S(t)
$$
and let $I_0$ be the desired number of infected at the equilibrium. 
The equations describing the interactions between removed and susceptible are not important now,
one has just to take into account that $S(t)$ decreases in time. Then, we can write
$$
\dot{E}=\gamma(t)I-\epsilon E, \ \dot{I}=\epsilon E - \delta I % \ \Rightarrow \ 
$$
that implies
$$
\begin{bmatrix}\dot{E} \cr \dot{I}\end{bmatrix}=\begin{bmatrix}-\epsilon & 0 \cr \epsilon & -\delta\end{bmatrix}\begin{bmatrix}E \cr I\end{bmatrix}+\begin{bmatrix}1 \cr 0\end{bmatrix}\begin{bmatrix}0 & 1\end{bmatrix}\begin{bmatrix}E \cr I\end{bmatrix}\gamma(t).
$$
This leads to the following correspondences with the matrices and the equilibrium point entering Theorem \ref{Main}
$$
F=\begin{bmatrix}-\epsilon & 0 \cr \epsilon & -\delta\end{bmatrix}, \ g=\begin{bmatrix}1 \cr 0\end{bmatrix}, \ H=\begin{bmatrix}0 & 1\end{bmatrix}, \ x_{eq}=\begin{bmatrix}\frac{\delta}{\epsilon} \cr 1\end{bmatrix}I_0, \ I_0>0,
$$
where the polynomial defining the desired system dynamics is given by $\Delta(s)=s+\lambda$ with $\lambda>0$.
We can alternate lockdown and freedom periods where, for simplicity, we assume that the contact rate is equal to $\beta_L$ and $\beta_F$, respectively.
It is easy to see that the couple $(F,g)$ is controllable while the condition on the product of the determinants becomes
$$
\det[F+gH\gamma_1^{(2)}]\det[F+gH\gamma_2^{(1)}]=\epsilon^2\big(\delta-\gamma_1^{(2)}\big)\big(\delta-\gamma_2^{(1)}\big)<0.
$$
This latter permits to conclude that sliding-mode SEIR control works properly as long as
\begin{equation}\label{CNS_SEIR}
\gamma_1(t):=\beta_LS(t)<\delta<\beta_FS(t):=\gamma_2(t).
\end{equation}
Since $Hx_{eq}=I_0>0$, from Theorem \ref{Main} one also obtains that the lockdown has to be introduced 
when $K(x-x_{eq})<0$. Here, exploiting the arguments of the theorem's proof contained in Appendix,  
the matrix $K$ is calculated using the controllability matrices ${\mathcal R}$ and ${\mathcal R}_c$
of the original system and of that in controllability canonical form \cite{Kailath79}, respectively. In particular, 
to obtain $\Delta(s)=s+\lambda$ one has 
$$
K_c=\begin{bmatrix}-\lambda & -1\end{bmatrix} \ \Rightarrow \ K=K_c{\mathcal R}_c{\mathcal R}^{-1}=\begin{bmatrix}-1 & \frac{\delta-\lambda}{\epsilon}\end{bmatrix}. 
$$
Hence, one obtains
$$
\mbox{Lockdown} \ \Leftrightarrow \ \begin{bmatrix}-1 & \frac{\delta-\lambda}{\epsilon}\end{bmatrix}\begin{bmatrix}E-E_0 \cr I-I_0\end{bmatrix}<0,
$$
a condition that, using $E_0=\frac{\delta}{\epsilon}I_0$, can be rewritten as
$$
\mbox{Lockdown} \ \Leftrightarrow \ \epsilon E -\delta I +\lambda(I-I_0)>0 \ 
%\Leftrightarrow \ \epsilon E +(\lambda-\delta)I>\lambda I_0 
\ \Leftrightarrow \ \lambda(I-I_0)+\dot{I}>0.
$$
We have so found the form of the sliding-mode controller anticipated in the description of Fig. \ref{Fig1} 
that, remarkably, is completely independent of the system parameters.\\
%%%%%%%%%%%%%
As said, the sliding surface will be reached and maintained only if \eqref{CNS_SEIR} holds true.
We can now reconsider the numerical experiments in Fig. \ref{Fig1}, 
where any state trajectory is generated using the SEIR with $\delta=0.2$.
When $\beta_L=0.2$, the condition in \eqref{CNS_SEIR} is satisfied since $S(0)<1$. 
This explains why all the associated trajectories (solid lines) quickly converge to the sliding
surface.\\
%that corresponds to having $\det[F+gH\gamma_1^{(2)}]\det[F+gH\gamma_2^{(1)}]<0$.
The two situations where \eqref{CNS_SEIR} is not satisfied are 
\begin{itemize}
\item $\beta_LS(t)>\delta$. In fact, %the sign of the two determinants is negative, 
one system eigenvalue is positive and the other one negative both
during the lockdown and in absence of restrictions. Hence, the epidemic grows independently of any control action (even during the lockdown)
until $S(t)$ becomes sufficiently small to satisfy the condition in \eqref{CNS_SEIR}.
This is exactly what happens in Fig. \ref{Fig1} when $\beta_L=0.21$ is adopted. 
Only when $S(t)$ decreases enough the associated trajectories (dashed lines) 
are attracted by the sliding line;
\item $\beta_FS(t)<\delta$. The two % the two determinants are both positive, 
system eigenvalues are now real and negative both 
during the lockdown and the freedom period. The sliding surface can not be reached and maintained because the epidemic dies down on its own, independently of any action. %basta restare sempre in FREE
\end{itemize}
Note also that, when \eqref{CNS_SEIR} is satisfied, the feedback law makes the equilibrium point 
$I_0>0$ asymptotically stable but then, when $S(t)$ becomes small enough, the second case described above 
will take place. The system will leave the sliding surface and the epidemic will end without the need of any restriction.\\
As a final but important note, Theorem \ref{Main} provides the necessary and sufficient conditions for asymptotic stability without % open neighborhood of of $x_{eq}$ 
specifying the domain of attraction ${\mathcal I}$ around the equilibrium point $x_{eq}$. This will depend on the particular system under study.
Remarkably, \emph{in the SEIR case also global convergence holds}, i.e. convergence is ensured
for any non null system initial condition if \eqref{CNS_SEIR} is fullfilled. The proof of this result is somewhat technical and
can be found in Appendix, see section \ref{Global}.

\subsubsection{SAIR} Using the \emph{SAIR}, infected are divided into two classes,
denoted by $A$ and $I$. Class $A(t)$ contains asymptomatic or paucisymptomatic who 
 can either directly recover with a rate established by $\epsilon_2$ or 
move to the second class $I(t)$ with a rate $\epsilon_1$. From $I(t)$, they can then recover
 with a rate $\delta$.  Dynamics are thus given by
%This leads to the following \emph{SAIR model} %which we call SIIR %modified SEIR   
 \begin{subequations}\label{SAIReqMT}
\begin{align}
\dot{S}(t)&=- \beta(t)S(t)\big(A(t)+I(t)\big) \\% \ a>0; \\
\dot{A}(t)&= \beta(t)S(t)\big(A(t)+I(t)\big) - \big(\epsilon_1+\epsilon_2\big)A(t) \\
\dot{I}(t)&= \epsilon_1A(t)-\delta I(t) \\ %, \ b>0; \\
\dot{R}(t)&=\epsilon_2A(t)+\delta I(t). %\\
%y(t)&=&\frac{1}{H}I(t). 
\end{align}
\end{subequations}
The infection rate $\beta(t)$ now describes how the interaction between susceptible $S(t)$
and the two classes $A(t),I(t)$ of infected evolves in time.\\ 
To exploit Theorem \ref{Main}, by adopting arguments very similar to those introduced
in the SEIR case, the following matrices and equilibrium point are derived 
$$
F=\begin{bmatrix}-(\epsilon_1+\epsilon_2) & 0 \cr \epsilon_1 & -\delta\end{bmatrix}, \ g=\begin{bmatrix}1 \cr 0\end{bmatrix},
$$ 
$$
H=\begin{bmatrix}1 & 1\end{bmatrix}, \ x_{eq}=\begin{bmatrix}\frac{\delta}{\epsilon_1} \cr 1\end{bmatrix}I_0, \ I_0>0.
$$
As done before, the contact rate may switch between the levels  $\beta_L$ and $\beta_F$
during the lockdown and the freedom period, respectively.
Also in this case the couple $(F,g)$ is controllable while the condition on the determinants becomes
$$
\det[F+gH\gamma_1^{(2)}]\det[F+gH\gamma_2^{(1)}]
$$
$$
=\big(\delta(\epsilon_1+\epsilon_2)-\gamma_1^{(2)}(\delta+\epsilon_1)\big)\big(\delta(\epsilon_1+\epsilon_2)-\gamma_2^{(1)}(\delta+\epsilon_1)\big)<0.
$$
Hence, the key condition for effectiveness of sliding-mode SAIR control is
\begin{equation}\label{CNS_SAIR}
\gamma_1(t):=\beta_LS(t)<\frac{\delta(\epsilon_1+\epsilon_2)}{\delta+\epsilon_1}<\beta_FS(t):=\gamma_2(t).
\end{equation}
One can thus see how the dynamics of asymptomatic people influence the
control threshold through the parameters $\epsilon_1,\epsilon_2$.
The desired system dynamics are still defined by $\Delta(s)=s+\lambda$ with $\lambda>0$.
So, one has $K_c=\begin{bmatrix}-\lambda & -1\end{bmatrix}$ and using again 
$K=K_c{\mathcal R}_c{\mathcal R}^{-1}$, simple calculations lead to
$$
K=\begin{bmatrix}-1 & \frac{\delta-\lambda}{\epsilon_1}\end{bmatrix}.
$$ 
This implies the following control law
$$
\mbox{Lockdown} \ \Leftrightarrow \ \lambda(I-I_0)+\dot{I}>0
$$
that coincides with that achieved in the SEIR case. It is also easy to see that the sliding surface now satisfies the equation
$$
\epsilon_1 A+(\lambda-\gamma) I = I_0 \lambda.
$$
Remarkably, \emph{even in the SAIR case global convergence holds}, 
see section \ref{Global} in Appendix.

\subsubsection{SEAIR}

The SEAIR model is a generalization of the SEIR and SAIR 
that embeds both the exposed and the asymptomatic class. It is given by
 \begin{subequations}%\label{SEIReq}
\begin{align}
\dot{S}&=-\beta(t)S(t)(A+I)\\ 
\dot{E}&=\beta(t)S(t)(A+I)-\epsilon E\\
\dot{A}&=\epsilon E-(\epsilon_1+\epsilon_2)A\\
\dot{I}&=\epsilon_1A-\delta I\\
 \dot{R}&=\epsilon_2A+\delta I.
\label{SEAIR}
\end{align}
\end{subequations}
Using the same arguments introduced for the analysis of the SEIR and SAIR,
we can focus on the variables $(E,A,I)$ and use $\gamma(t)$ to account for the other hidden nonlinear dynamics.
One obtains the matrices
$$
F=\begin{bmatrix}-\epsilon & 0 & 0 \cr \epsilon & -(\epsilon_1+\epsilon_2) & 0 \cr 0 & \epsilon_1 & -\delta\end{bmatrix}, \ g=\begin{bmatrix}1 \cr 0 \cr 0\end{bmatrix},
$$
$$
H=\begin{bmatrix}0 & 1 & 1\end{bmatrix}, \ x_{eq}=\begin{bmatrix}\frac{\delta(\epsilon_1+\epsilon_2)}{\epsilon\epsilon_1} \cr \frac{\delta}{\epsilon_1} \cr 1\end{bmatrix}I_0, \ I_0>0.
$$
We still assume that the contact rate can switch only between $\beta_L$ and $\beta_F$
alternating lockdown and freedom periods, respectively.
Similarly to the other cases, the couple $(F,g)$ is controllable while one now has %condition is %on the product of the determinants becomes
$$
\det[F+gH\gamma_1^{(2)}]\det[F+gH\gamma_2^{(1)}]
$$
$$
=\epsilon^2\big(\delta(\epsilon_1+\epsilon_2)-\gamma_1^{(2)}(\delta+\epsilon_1)\big)\big(\delta(\epsilon_1+\epsilon_2)-\gamma_2^{(1)}(\delta+\epsilon_1)\big)<0.
$$
Thus, sliding-mode SEAIR control requires fulfilment of the same condition 
related to the SAIR and reported in \eqref{CNS_SAIR}, i.e. once again one needs
\begin{equation}\label{CNS_SEAIR}
\gamma_1(t):=\beta_LS(t)<\frac{\delta(\epsilon_1+\epsilon_2)}{\delta+\epsilon_1}<\beta_FS(t):=\gamma_2(t).
\end{equation}
Let now the Hurwitz polynomial be $\Delta(s)=s^2+\Delta_1s+\Delta_0$.
To derive the sliding-mode control form,
calculations are more difficult than in the previous cases so, to make the exposition more compact, we think backwards
verifying that 
\begin{equation}\label{SEAIRclaw}
\mbox{Lockdown} \ \Leftrightarrow \ddot{I}+\Delta_1\dot{I}+\Delta_0(I-I_0)>0.
\end{equation}
In fact, in view of the definition of $\Delta(s)$, 
one now has 
\begin{equation}\label{KcSEAIR}
K_c=\begin{bmatrix}-\Delta_0 & -\Delta_1 & -1\end{bmatrix}.
\end{equation}
In addition, 
$$
\dot{I}=\epsilon_1 A-\delta I \ \Rightarrow \ \ddot{I}=\epsilon_1 \dot{A}-\delta \dot{I}=\epsilon_1[\epsilon E -(\epsilon_1+\epsilon_2)A]-\delta[\epsilon_1 A-\delta I]
$$
that implies 
$$
\ddot{I}+\Delta_1\dot{I}+\Delta(I-I_0)=\epsilon_1 \dot{A}-\delta \dot{I}
$$
$$
=\epsilon_1[\epsilon E -(\epsilon_1+\epsilon_2)A]-\delta[\epsilon_1 A-\delta I]+\Delta_1[\epsilon_1 A-\delta I]+\Delta_0(I-I_0).
$$
This means that the lockdown's condition $\ddot{I}+\Delta_1\dot{I}+\Delta_0(I-I_0)>0$ can be rewritten as
$$
\epsilon\epsilon_1E-\epsilon_1(\epsilon_1+\epsilon_2+\delta-\Delta_1)A+(\delta^2-\delta\Delta_1+\Delta_0)I-\Delta_0I_0>0.
$$
Letting
\begin{equation}\label{KSEAIR}
K:=h\begin{bmatrix}-\epsilon\epsilon_1 & \epsilon_1(\epsilon_1+\epsilon_2+\delta-\Delta_1) & -(\delta^2-\delta\Delta_1+\Delta_0)\end{bmatrix}, %\ 
\end{equation}
with $h$ being any positive scale factor,
the previous condition becomes
$$
-Kx-h\Delta_0I>0 \ \Rightarrow \ Kx+h\Delta_0I_0<0 \ \Rightarrow \ K(x-x_{eq})<0
$$
(since $Kx_{eq}=-h\Delta_0I_0$). To derive $h$, we know that 
the relationship $K{\mathcal R}=K_c{\mathcal R}_c$ must hold where, in the SEAIR case,
the controllability matrices are
%$K=K_c{\mathcal R}_c{\mathcal R}^{-1} \ \Leftrightarrow \ K{\mathcal R}=K_c{\mathcal R}_c
%$$
%where
$$
{\mathcal R}=\begin{bmatrix}1 & -\epsilon & \epsilon^2 \cr 0 & \epsilon & -\epsilon(\epsilon+\epsilon_1+\epsilon_2) \cr 0 & 0 & \epsilon\epsilon_1\end{bmatrix},  
$$
$$\tiny
{\mathcal R}_c=\begin{bmatrix}0 & 0 & 1 \cr 0 & 1 & -(\delta+\epsilon+\epsilon_1+\epsilon_2) \cr 1 & -(\delta+\epsilon+\epsilon_1+\epsilon_2) & -\epsilon\delta-(\epsilon+\delta)(\epsilon_1+\epsilon_2)+(\delta+\epsilon+\epsilon_1+\epsilon_2)^2\end{bmatrix}.
$$
%This, together with the already stated correspondences $$
%K_c=\begin{bmatrix}-\Delta_0 & -\Delta_1 & -1\end{bmatrix}, \ K=h\begin{bmatrix}-\epsilon\epsilon_1 & \epsilon_1(\epsilon_1+\epsilon_2+\delta-\Delta_1) & -(\delta^2-\delta\Delta_1+\Delta_0)\end{bmatrix},
%$$
Recalling \eqref{KcSEAIR} and \eqref{KSEAIR}, it becomes easy to conclude that $h=\frac{1}{\epsilon\epsilon_1}$ is necessary and also sufficient for  $K{\mathcal R}=K_c{\mathcal R}_c$ to hold.
Thus, 
$$
K=\begin{bmatrix}-1 & \frac{\epsilon_1+\epsilon_2+\delta-\Delta_1}{\epsilon} & -\frac{\delta^2-\delta\Delta_1+\Delta_0}{\epsilon\epsilon_1}\end{bmatrix}
$$
is  the unique $K$ obtainable from Theorem \ref{Main}, hence proving the correctness of the % and indeed leads to the 
postulated feedback law in \eqref{SEAIRclaw} that, as in the previous cases, does not depend on system parameters.

%{\color{blue}
%\subsubsection{How much asymptomatic people can make control harder}
%
%}

%deducibile dal TH, che conduce alla ben nota legge 
%$$
%\mbox{LOCK} \ \Leftrightarrow \ \ddot{I}+\Delta_1\dot{I}+\Delta_0(I-I_0)>0, \ \mbox{con} \ \Delta(s)=s^2+\Delta_1s+\Delta_0 \ \mbox{polinomio stabile}
%$$

%
%{\color{red}
%\section{Modello generale}
%
%Il procedimento \`e ovviamente applicabile a qualsiasi altro modello compartimentale similare, fermo restando che
%\begin{itemize}
%\item il procedimento permette di determinare quell'UNICA $K$ che risolve il problema (in realt\`a infinite $K$, una per ogni polinomio stabile)
%\item risolvere il problema significa che la convergenza LOCALE allo sliding \`e garantita, non quello GLOBALE (eccetto per i modelli 2D SAIR/SEIR)
%\item non \`e neppure detto, in generale, che la legge di controllo DALLO STATO sia esprimibile in termini solo di $I$ e delle sue derivate, il che lo renderebbe (probabilmente sempre) insensibile ai parametri
%\end{itemize}
%}

 \section{Conclusions}
 
 The convergence of sliding-mode controllers subject to strong structural constraints on the feedback law  has been investigated in this paper. 
The fundamental assumption that permeates this work is that the control input may assume values only over two bounded and
 disjoint sets that could be also non perfectly known. 
 The necessary and sufficient conditions for an effective control of a class of time-varying continuous-time linear systems,
that includes all the time-invariant linear state-space models, 
 have been obtained. 
A rather general family of controllers is so derived which guarantees to reach the sliding surface and maintain the desired
equilibrium point. Notably, the analysis is only apparently restricted to the linear setting since 
the class of dynamic systems here introduced is wider than expected.
In fact, the cases 
reported in the previous section illustrate how the time-varying component
present in our model can be conveniently used to describe hidden nonlinear dynamics. %allows also 
This permits to gain fundamental insights also on control of a wide class of nonlinear systems, % e.g. during an epidemic. 
making our findings relevant for the most varied applications. % of our findings are so potentially infinite. % and %In particular, this family includes several
We have provided examples that regard well known epidemiological models
 that incorporate the classes
of exposed and asymptomatic people, especially important e.g. to describe COVID-19 dynamics. 
Many previous works on epidemic control can thus be interpreted
from a broader perspective, becoming special cases of the sliding-mode theory under feedback constraints here developed.\\
By exploiting the time-varying component that is integrated in the class of dynamic systems here proposed,
SEIR and SAIR reduce to two-dimensional models.
The fact that global convergence of sliding-mode controllers can be proved for these two systems 
leads also to an interesting open problem. In particular, we conjecture that global convergence
holds for any positive and controllable two-dimensional system and we plan to investigate %in depth
such issue  in the next future.

\section{Appendix}

\subsection{Proof of Theorem \ref{Main}}

%\noindent In this section the proof of Theorem \ref{Main} is reported.
A simple preliminary lemma %(whose proof is reported for the sake of completeness) 
is first obtained.

\begin{lemma} \label{LemmaSM} Given
$$
\dot{x}=Fx+gu, \ F \ \mbox{asymptotically stable}, \ u(t) \ \mbox{bounded}, \ u_M:=\sup_t \ |u(t)|
$$
there exist a real number $a>0$ independent of $u_M$ and an open neighborhood ${\mathcal I}$ of $x=0$ such that %exist, with the following properties
\begin{itemize}%[leftmargin=1cm]
\item ${\mathcal I}$ is an invariant set, i.e. $x(0) \in {\mathcal I}$ implies $x(t) \in {\mathcal I}$ for any $t \ge 0$;
\item $\|{\mathcal I}\|:=sup_{x \in {\mathcal I}} \ \|x\| = au_M$.
\end{itemize}
\end{lemma}
%\hfill$\heartsuit$

\noindent {\bf Proof.} \ 
Let $P=P^T>0$ be the (unique) solution of the Lyapunov equation $F^TP+PF=-I$, and define
$$
{\mathcal I}(b):=\left\{ \ x \in {\mathbb R}^n: \ x^TPx < b \ \right\}, \ b>0 \ \Rightarrow \ \|{\mathcal I}(b)\|=\sqrt{\frac{b}{\lambda_{MIN}( P )}}
$$
Exploiting
$$
\frac{d}{dt} x^T(t)Px(t)=\dot{x}^TPx+x^TP\dot{x}
$$
$$
=[Fx+gu]^TPx+x^TP[Fx+gu]=x^T(F^TP+PF)x+2x^TPgu=
$$
$$
=-\|x(t)\|^2+2\|x(t)\|\frac{x^T(t)}{\|x(t)\|}Pgu(t) 
$$
$$
\le -\|x(t)\|^2+2\|x(t)\|u_M \max_{v: \ \|v\|=1} \ v^TPg=
$$
$$
=-\|x(t)\| \Big(  \|x(t)\|-2u_M \max_{v: \ \|v\|=1} \ v^TPg \Big)
$$
and
$$
\|x(t)\| \ge \sqrt{\frac{x^T(t)Px(t)}{\lambda_{MAX}( P )}}
$$
it follows that
$$
x^T(t)Px(t) > 4\lambda_{MAX}( P )u_M^2\Big( \max_{v: \ \|v\|=1} \ v^TPg \Big)^2
$$
implies
$$
\frac{d}{dt} x^T(t)Px(t) < 0.
$$
Let $b>0$ satisfy
$$
b>4\lambda_{MAX}( P )u_M^2 \Big( \max_{v: \ \|v\|=1} \ v^TPg \Big)^2. 
$$
For a suitable $\mu>1$, one then has 
%\ \Rightarrow \ 
$$
\|{\mathcal I}(b)\|=2\mu \Big( \max_{v: \ \|v\|=1} \ v^TPg \Big) \sqrt{\frac{\lambda_{MAX}( P )}{\lambda_{MIN}( P )}}u_M:= au_M, 
$$
with % $a$ defined as
$$
a:=2\mu \Big( \max_{v: \ \|v\|=1} \ v^TPg \Big) \sqrt{\frac{\lambda_{MAX}( P )}{\lambda_{MIN}( P )}}.
$$
Now, we want to show that ${\mathcal I}={\mathcal I}(b)$ satisfies the theorem statement by proving its invariance. In fact, assuming $x^T(0)Px(0)<b$, it holds that $x^T(t)Px(t)<b$ for any $t\ge 0$, otherwise a time instant $T > 0$ would exist such that $x^T(T)Px(T) \ge b$. Because of the trajectory continuity, the set ${\mathcal Z}:=\{ \ t \ge 0: \ x^T(t)Px(t) = b \ \}$ would be non-empty. Also, it would be equipped with an inferior $\bar{t}:=inf {\mathcal Z}$ which would imply $x^T(t)Px(t) < b$ for any $0<t<\bar{t}$, while $x^T(\bar{t})Px(\bar{t}) = b$. One would thus obtain that $\frac{d}{dt} x^T(t)Px(t)$ would be strictly less than zero in a left neighborhood of $\bar{t}$, hence making $x^T(t)Px(t) > x^T(\bar{t})Px(\bar{t})$ for some $t<\bar{t}$. This contradicts the inferior property of $\bar{t}$. So, it must hold that ${\mathcal Z}=\emptyset$, from which $x^T(t)Px(t) < b$ for any $t\ge 0$ that is equivalent to the invariance property. 
%$\qed$ %\hfill$\heartsuit$
{\flushright$\Box$}\\ 

\noindent We are now in a position to prove the theorem. This will be done in five steps.
The first four prove the sufficiency of the two conditions regarding controllability and the product of
determinants while  the last step discusses their necessity. \\

\noindent {\bf Step 1 [Existence of $x_{eq}, \gamma_0$ and equations in terms of $x-x_{eq}$]} \\
 Let
$$
\Delta_F(s)=s^n+\dots+a_1s+a_0
$$
be the characteristic polynomial of $F$. Consider the system in controllability canonical form (where also $H$ has been accordingly modified)
$$
T^{-1}FT=F_c, \ T^{-1}g=g_c, \ H_c:=HT,
$$
with $T={\mathcal R}{\mathcal R}_c^{-1}$ where ${\mathcal R}$ and ${\mathcal R}_c$
are the controllability matrices of the original system and of that in controllability canonical form \cite{Kailath79}, respectively. 
The structure of these matrices permits to easily verify that  
$$
\det[F_c+g_cH_c\gamma(t)]=(-1)^n[a_0-H_ce_1\gamma(t)].
$$
So, the second assumption implies
$$
[a_0-H_ce_1\gamma_1^{(2)}][a_0-H_ce_1\gamma_2^{(1)}]<0
$$
which means that $a_0-H_ce_1\gamma$ is linear in $\gamma$ and assumes opposite signs at $\gamma_1^{(2)},\gamma_2^{(1)}$.
This implies that there exists a unique value $\gamma_0$, necessarily falling in the interval $(\gamma_1^{(2)}, \gamma_2^{(1)})$,  such that
$$
\det[F_c+g_cH_c\gamma_0]=0 \ \Leftrightarrow \ \det[F+gH\gamma_0]=0.
$$
Also, this shows that a point $x_{eq} \ne 0$ exists such that $[F+gH\gamma_0]x_{eq}=0$ (and one easily sees that $\mbox{rank}[F_c+g_cH_c\gamma_0]=n-1$, so that $x_{eq}$ is uniquely determined except for a multiplicative constant), which implies $Fx_{eq}=-gH\gamma_0x_{eq}$. 
Hence
$$
\dot{x}=F(x-x_{eq})+Fx_{eq}+gu=F(x-x_{eq})+g[u-H\gamma_0x_{eq}]
$$
and, after defining $y:=x-x_{eq}$ and $v:=u-H\gamma_0x_{eq}$, one has
$$
\dot{y}=Fy+gv
$$
\noindent {\bf Step 2 [Construction of the sliding surface and related equations]}\\ 
Consider again the controllability canonical form
$$
\dot{z}=F_cz+g_cv, \ z:=T^{-1}y=\begin{bmatrix}\bar{z} \cr z_n\end{bmatrix}
$$
where the last entry of $z$ has been highlighted. Consider any matrix $K_c:=\begin{bmatrix} k_1 & k_2 & \dots & k_{n-1} & -1\end{bmatrix}:=\begin{bmatrix}\bar{K} & -1\end{bmatrix}$ such that $\Delta(s):=s^{n-1}-k_{n-1}s^{n-2}+\dots-k_2s-k_1$ is an Hurwitz polynomial.
Then, define $K$ through $K=K_cT^{-1}$, the sliding surface as $Kx=Kx_{eq}$, and the following change of variables
$$
z=\begin{bmatrix}\bar{z} \cr z_n\end{bmatrix}=\begin{bmatrix}I & 0 \cr \bar{K} & -1\end{bmatrix}\begin{bmatrix}\bar{z} \cr K_cz\end{bmatrix}:=U\begin{bmatrix}\bar{z} \cr K_cz\end{bmatrix}, \ U^{-1}=U, \ w:=\begin{bmatrix}\bar{z} \cr K_cz.
\end{bmatrix}
$$
First, notice that
$$
\dot{z}=F_cz+g_cv \ \Rightarrow \ K_c\dot{z}=K_cF_cz+K_cg_cv=K_cF_cz-v %\ 
$$
$$
\Rightarrow \ K_c\dot{z}=r:=K_cF_cz-v \ \Rightarrow \ v=K_cF_cz-r
$$
so that we can express the differential equations in terms of the new input $r$ as follows
$$
\dot{z}=F_cz+g_cK_cF_cz-g_cr=(I+g_cK_c)F_cz-g_cr.
$$
This implies
$$
\dot{w}=U(I+g_cK_c)F_cUw-Ug_cr=\begin{bmatrix}M & -\bar{g}_c \cr 0 & 0\end{bmatrix}w+\begin{bmatrix}0 \cr 1\end{bmatrix}r
$$
where $M$ is in companion form with the last row equal to $\bar{K}$, so that $\Delta_M(s)=\Delta(s)$ 
and $\bar{g}_c$ is like $g_c$ in $(n-1)$ dimensions.
Also, by recalling the definition of $w$ and the relationships between $u,v,r$, it follows that
\begin{eqnarray}
\dot{\bar{z}}&=&M\bar{z}-\bar{g}_c(K_cz) \cr
(K_c\dot{z})&=&r=K_cF_cz+Hx_{eq}\gamma_0-u
\label{CACCA}
\end{eqnarray}

\noindent {\bf Step 3 [Condition for sliding activation]} \\
By exploiting the second part of
\eqref{CACCA} one obtains
$$
K_c\dot{z}=K_cF_cz+Hx_{eq}\gamma_0-u=K_cF_cz+Hx_{eq}\gamma_0-Hx\gamma(t)
$$
$$
=K_cF_cz+Hx_{eq}\gamma_0-Hy\gamma(t)-Hx_{eq}\gamma(t)=
$$
$$
=[K_cF_c-H_c\gamma(t)]z+Hx_{eq}[\gamma_0-\gamma(t)]
$$
where $Hx_{eq} \ne 0$. Otherwise $(F+gH\gamma_0)x_{eq}=0$ would imply $Fx_{eq}=0$ and also $(F+gH\gamma_1^{(2)})x_{eq}=0$, which would contradict the invertibility of the last matrix (recall that $x_{eq}\ne 0$). Since $K_cz=KTz=Ky=K(x-x_{eq})$, it follows that 
$$
K\dot{x}=[KF-H\gamma(t)](x-x_{eq})+Hx_{eq}[\gamma_0-\gamma(t)]
$$
and, by defining
$$
m:=\min\big(|Hx_{eq}(\gamma_0-\gamma_1^{(2)})|, |Hx_{eq}(\gamma_0-\gamma_2^{(1)})| \big)
$$
and resorting to the control law described in the theorem statement, in case of
\begin{equation}
|[KF-H\gamma(t)](x-x_{eq})|<m-c, \ m>c>0
\label{ACC}
\end{equation}
it holds that
$$
|K\dot{x}|<c \ \ \mbox{and} \ \  \mbox{sign}[K\dot{x}]\mbox{sign}[K(x-x_{eq})]<0.
$$
This implies that $K[x(t)-x_{eq}]=0$ from some $t <\frac{K[x(0)-x_{eq}]}{c}$ onwards, if \eqref{ACC} is satisfied by $x(t)$ for all $t \ge 0$, so establishing the sliding phenomenon in the surface $Kx=Kx_{eq}$. This fact is surely verified if
$$
\|x(t)-x_{eq}\|<\frac{m-c}{\mu}, \ \forall t \ge 0,
$$
with
 %\ \mbox{with} \ 
 $$
 \mu:=\max_{(v,\gamma): \ \|v\|=1, \gamma_1 \in I_1, \gamma_2 \in I_2} \ \{ \ |[KF-H\gamma_1]v|, \ |[KF-H\gamma_2 ]v| \ \}.
$$
Moreover, recalling that $w=Uz$ and $z=T^{-1}(x-x_{eq})$, the previous condition can be rewritten as
$$
\|w(t)\| \le b \|x(t)-x_{eq}\|<\frac{m-c}{b\mu}:=A, \ \forall t \ge 0, \ \mbox{for some $b>0$}.
$$
This means that there exists a suitable $A>0$ such that
\begin{equation} \label{SLID}
\|w(t)\| < A, \ \forall t \ge 0 % \ \Rightarrow \ 
\end{equation}
and this implies that % implies
$$
\exists \ \bar{t}>0: \ |K[x(t)-x_{eq}]| \ \mbox{is monotone decreasing} 
$$
and
$$
K[x(t)-x_{eq}]=0, \ \forall t\ge \bar{t},
$$
so leading to the sliding establishment.
\medskip

\noindent {\bf Step 4 [The invariant neighborhood of $x_{eq}$]} \\
By applying Lemma \ref{LemmaSM} to the first part of \eqref{CACCA}, rewritten as
$$
\dot{\bar{w}}=M\bar{w}-\bar{g}_cw_n, \ w:=\begin{bmatrix}\bar{w}^T & w_n\end{bmatrix}^T,
$$
assuming that $w_n$ evolves in $\left(-d, d \right)$ implies the existence of an open neighborhood ${\mathcal I}_{\bar{w}}$ of $\bar{w}=0$ which is invariant. Moreover, for some $a>0$ one has
$$
\|\bar{w}(t)\|<ad, \ \forall t\ge 0.
$$
It follows that we can build the open neighborhood ${\mathcal I}_w:={\mathcal I}_{\bar{w}} \times \left(-d, d\right)$ of $w=0$ 
which is invariant and such that 
$$
\|w(t)\|=\sqrt{\|\bar{w}(t)\|^2+w_n^2}< d \sqrt{a^2+1} <A, \ \forall t \ge 0
$$
if $d$ satisfies % is chosen in such a way that
$$
0<d<\frac{A}{\sqrt{a^2+1}}.
$$
Indeed, recalling also \eqref{CACCA}, this choice of $d$ ensures that $w_n$ evolves in $\left(-d, d \right)$, which is exactly what was required for guaranteeing both the invariance of ${\mathcal I}_w$ and the convergence to zero of $w_n(t)=K[x(t)-x_{eq}]$. Once the surface $w_n=0$ is reached,  system dynamics on the sliding surface are simply given by $\dot{\bar{w}}=M\bar{w}$. This leads to a trajectory contained in ${\mathcal I}_{\bar{w}}$ (the input is now zero) in $(n-1)$ dimensions and in ${\mathcal I}_{\bar{w}} \times \{ \ 0 \ \}$ (which is a subset of ${\mathcal I}_w$) in $n$ dimensions.
Hence, the invariance is maintained also after reaching the sliding surface and $w$ tends to zero without exiting from both $w_n=0$ and ${\mathcal I}_w$.
\noindent As far as $x(t)-x_{eq}$ is concerned, the linear relationship between it and $w$ leads to a corresponding open neighborhood ${\mathcal I}$ of $x_{eq}$ %in the state variables $x$, 
that enjoys all the properties required in the theorem statement.\\

\noindent {\bf Step 5 [Necessity of the condition regarding controllability and negative sign of determinants product]} \\
In the previous four steps, sufficiency of the sliding-mode control conditions regarding controllability and sign of the determinants product has been proved. We now demonstrate their necessity to conclude the theorem. 
 If $(F,g)$ is not a controllable pair, any uncontrollable initial condition in $\mathcal{I}$ cannot be driven to any desired state, hence sliding-mode control could not work. The same conclusion can be obtained if the condition on the determinants is violated  through the arguments developed in Step 1. 
 In fact, one could not find any $\gamma_0 \in (\gamma_1^{(2)},\gamma_2^{(1)})$ and corresponding $x_{eq}\ne 0$, as 
 $\det[F+gH\gamma]$ linearly depends on $\gamma$.

\subsection{Global convergence of sliding-mode under feedback constraints for SEIR and SAIR}\label{Global}

In the main part of this paper we have provided the necessary and sufficient conditions for asymptotic stability of a class of time-varying systems
under sliding-mode control subject to feedback constraints.
The theorem guarantees the existence of a domain of attraction ${\mathcal I}$ around the equilibrium point $x_{eq}$. 
The shape of ${\mathcal I}$ will depend on the particular system under study.
As anticipated, in the SEIR and SAIR case global convergence holds, i.e. convergence is ensured
with  ${\mathcal I}$ set to the entire state space excluding the origin.
The proof of this result is reported below.
We just consider the SEIR case since the arguments concerning SAIR
are completely analogous.  

\subsubsection{Switching evolution}

Let's define the region ${\mathcal R}_{FREE}$ as the region where $\gamma(t)$ assumes the value $\gamma_2:=\gamma_F$
$$
{\mathcal R}_{FREE}:=\{ \ (E,I)\ge 0: \ \epsilon E+(\lambda-\delta)I<\lambda I_0 \ \}
$$
and, analogously, the complementary region ${\mathcal R}_{LOCK}$ corresponding to $\gamma_1:=\gamma_L$.
So, ${\mathcal R}_{FREE} \cup {\mathcal R}_{LOCK} \cup {\mathcal L}={\mathbb R}_+^2$, where ${\mathcal L}$ is the sliding line included in the positive quadrant. It is easy to see that a whole neighborhood of the origin is completely included in ${\mathcal R}_{FREE}$, and that ${\mathcal L}$ is either a segment or an half straight line (depending on the value of $\lambda>0$). Let's introduce the Frobenius's eigenvalue $\lambda_{FR}>0$ associated with the matrix $F_{FREE}:=F+gH\gamma_F$ and the corresponding positive eigenvector $w_{FR}$. Simple computations lead to
$$
w_{FR}=2\begin{bmatrix}\lambda_{FR}+\delta & \epsilon\end{bmatrix}^T=\begin{bmatrix}(\delta-\epsilon)+\sqrt{(\delta-\epsilon)^2+4\epsilon\gamma_{F}} & 2\epsilon\end{bmatrix}^T,   
$$
$$
\lambda_{FR}=\frac{-(\delta+\epsilon)+\sqrt{(\delta-\epsilon)^2+4\epsilon\gamma_{F}}}{2},
$$
so that the intersection of ${\mathcal L}$ with the half straight line $x=aw_{FR}, \ a \ge 0$ is easily found
$$
(E,I)=\begin{pmatrix}\frac{\lambda(\delta+\lambda_{FR})}{\epsilon(\lambda+\lambda_{FR})} & \frac{\lambda}{\lambda+\lambda_{FR}} \end{pmatrix}I_0>0.
$$ 
Since $\lambda,\lambda_{FR}>0$, the point $I_{FR}$ where the intersection takes place satisfies $0<I_{FR}<I_0$. This means that, regardless of the nature (finite or infinite) of ${\mathcal L}$, there always exists an intersection in the positive quadrant. But this implies that, starting from any $x(0) \in {\mathcal R}_{FREE}, \ x(0) \ne 0$, at a certain instant  $x(t)$ leaves ${\mathcal R}_{FREE}$ and touches the sliding line. This happens because, by using $w$ to denote the other eigenvector of $F_{FREE}$, $x(0)\ge 0, \ x(0)\ne 0$ implies $x(0)=aw_{FR}+bw$ (with $a>0$ in view of the positive systems properties), so that $x(t)$ tends to grow as $e^{\lambda_{FR}t}$ and to follow the direction of $w_{FR}$. Existence of the intersection at $I_{FR}$ also ensures the existence of a time instant $t>0$ such that $x(t)$ belongs to ${\mathcal L}$, hence leaving ${\mathcal R}_{FREE}$. Note that this is obvious in case of a finite sliding line, while in the infinite case the proof of existence of the aforementioned intersection is required. The same holds true for ${\mathcal R}_{LOCK}$ too. For any $x(0) \in {\mathcal R}_{LOCK}$, sooner or later ${\mathcal L}$ is reached again, in this case because of the asymptotic stability of $F+gH\gamma_L:=F_{LOCK}$ which makes $x(t)$ convergent to zero. So, it suffices to recall that a whole neighborhood of the origin is included in ${\mathcal R}_{FREE}$. This property of the sliding line allows us to study the sliding establishment by considering only initial conditions in ${\mathcal L}$.

\subsubsection{The sliding establishment zone}

Define
$$
I_{MAX}:=\frac{\lambda}{\lambda-\delta}I_0 \ \mbox{(if $\lambda>\delta$)}, \ I_{MAX}:=+\infty \ \mbox{(if $0<\lambda\le\delta$)}
$$
so that $I$, on the sliding line, can assume only values in $[0,I_{MAX}]$. Now we investigate the behavior of the sliding line points, by showing that
only the following three cases may arise: % three possibilities are allowed
\begin{itemize}
\item the flow goes from ${\mathcal R}_{FREE}$ towards ${\mathcal R}_{LOCK}$;
\item the flow goes from ${\mathcal R}_{LOCK}$ towards ${\mathcal R}_{FREE}$;
\item the flow goes both from ${\mathcal R}_{FREE}$ towards ${\mathcal R}_{LOCK}$ and from ${\mathcal R}_{LOCK}$ towards ${\mathcal R}_{FREE}$.
\end{itemize}
In the first two cases, when considered as mutually exclusive, the trajectory crosses the sliding line.
The third case corresponds to points belonging to the subset of ${\mathcal L}$ where the trajectory cannot escape from ${\mathcal L}$ itself, so establishing the sliding mode converging to $I_0$. It thus corresponds to the overlapping of the first two cases.

\subsubsection{Flow from ${\mathcal R}_{FREE}$ to ${\mathcal R}_{LOCK}$}

We start by analyzing which points are related to the first situation. This requires to assume $K\dot{x}<0$, with $\dot{x}=F_{FREE}x$, together with $K(x-x_{eq})=0$. We easily obtain the following inequality
$$
[\lambda^2-(\delta+\epsilon)\lambda-\epsilon(\gamma_{F}-\delta)]I<\lambda(\lambda-\epsilon-\delta)I_0
$$
which, for suitable $\lambda_1,\lambda_2>0$, can be rewritten as 
$$
(\lambda-\lambda_1)(\lambda+\lambda_2)I<\lambda(\lambda-\epsilon-\delta)I_0, \ \lambda_1>\epsilon+\delta.
$$
Obviously, both $I_0$ and $I_{FR}$ always satisfy the inequality. By defining
%(the first one because it belongs to the sliding establishment subset, the second one because of being the flow directed like $w_{FR}$). By defining
$$
I_1:=\frac{\lambda(\lambda-\epsilon-\delta)}{\lambda^2-(\delta+\epsilon)\lambda-\epsilon(\gamma_{F}-\delta)}
$$
the previous inequality becomes something like either $I>I_1$ or $I<I_1$, depending on the signs of the left and right side terms. 
One has also to take into account that $I$ cannot be either negative or exceed $I_{MAX}$. After discriminating among a certain number of cases, it follows that if $\lambda>\lambda_1$ the inequality's solution is given by $I<I_1$, with $I_1>I_0>I_{FR}$. However, $I_1>I_{MAX}$ if $\lambda_1<\lambda<\delta+\frac{\epsilon}{\delta}\gamma_{FR}$, so that 
\begin{itemize}
\item if $\lambda>\delta+\frac{\epsilon}{\delta}\gamma_{F}$, then $I<I_1$, with $I_{MAX}>I_1>I_0>I_{FR}>0$
\item if $\lambda_1<\lambda\le\delta+\frac{\epsilon}{\delta}\gamma_{F}$, then $I$ is any.
\end{itemize}
The second case is $\lambda_1\ge\lambda\ge\epsilon+\delta$, which leads again to $I$ being any, while the third case is $\epsilon+\delta>\lambda>0$, which leads to $I>I_1$, with $I_{MAX}>I_0>I_{FR}>I_1>0$. All of these outcomes are summarized below.\\
%\medskip

\noindent {\bf The flow goes from ${\mathcal R}_{FREE}$ to ${\mathcal R}_{LOCK}$}
\begin{itemize}
\item if $\lambda>\delta+\frac{\epsilon}{\delta}\gamma_{F}$, then $I<I_1$, with $I_{MAX}>I_1>I_0>I_{FR}>0$;
\item if $\delta+\epsilon\le\lambda\le\delta+\frac{\epsilon}{\delta}\gamma_{F}$, then $I$ is any;
\item if $\epsilon+\delta>\lambda>0$, then $I>I_1$, with $I_{MAX}>I_0>I_{FR}>I_1>0$.
\end{itemize}

\subsubsection{Flow from ${\mathcal R}_{LOCK}$ to ${\mathcal R}_{FREE}$}

Now we consider points related to the second situation. This requires to assume $K\dot{x}>0$, with $\dot{x}=F_{LOCK}x$, together with $K(x-x_{eq})=0$. The following inequality is easily obtained
$$
[\lambda^2-(\delta+\epsilon)\lambda+\epsilon(\delta-\gamma_L)]I>\lambda(\lambda-\epsilon-\delta)I_0.
$$
For suitable $\lambda_3,\lambda_4>0$ satisfying $0<\lambda_4<\delta<\lambda_3<\delta+\epsilon$, it
can be rewritten as 
$$
(\lambda-\lambda_3)(\lambda-\lambda_4)I>\lambda(\lambda-\epsilon-\delta)I_0.
$$
Again, $I_0$ always satisfies the inequality, and by defining
$$
I_2:=\frac{\lambda(\lambda-\epsilon-\delta)}{\lambda^2-(\delta+\epsilon)\lambda+\epsilon(\delta-\gamma_{L})}
$$
the previous inequality becomes something like either $I>I_2$ or $I<I_2$, depending on the signs of the left and right side terms. We have also to take into account that $I$ cannot be either negative or exceed $I_{MAX}$. %By distinguish among a certain number of cases, 
It follows that if $\lambda>\delta+\epsilon$ the inequality solution is given by $I>I_2$, with $I_{MAX}>I_0>I_2>0$. If 
 $\delta+\epsilon\ge\lambda\ge\lambda_3$ then $I$ can be any, while if $\delta<\lambda<\lambda_3$ it becomes $I<I_2$, with $I_2>I_0$. However, also in this case we have to verify whether $I_2<I_{MAX}$ and this actually holds only if $\delta<\lambda<\delta+\frac{\epsilon}{\delta}\gamma_{L}<\lambda_3$. Moreover, the same situation $I<I_2$, with $I_{MAX}>I_2>I_0$, occurs if $\delta\ge\lambda>\lambda_4$, while $I$ can assume
 any value if $0<\lambda<\lambda_4$. We summarize these outcomes below.\\

\noindent {\bf The flow goes from ${\mathcal R}_{LOCK}$ to ${\mathcal R}_{FREE}$}
\begin{itemize}
\item if $\lambda>\delta+\epsilon$, then $I>I_2$, with $I_{MAX}>I_0>I_2>0$;
\item if $\delta+\epsilon\ge\lambda\ge\delta+\frac{\epsilon}{\delta}\gamma_{L}$, then $I$ is any;
\item if $\lambda_4<\lambda<\delta+\frac{\epsilon}{\delta}\gamma_{L}$, then $I<I_2$, with $I_{MAX}>I_2>I_0>0$;
\item if $0<\lambda<\lambda_4$, then $I$ is any.
\end{itemize}

\subsubsection{The attractive sliding zone}

Considering all the results obtained so far, we can notice that, whenever the whole ${\mathcal L}$ is attractive at least in one direction (which means that all the trajectories go either from ${\mathcal R}_{LOCK}$ towards ${\mathcal R}_{FREE}$ or conversely), the trajectory can cross ${\mathcal L}$ at most once before falling into the sliding establishment. 
Therefore, it holds that % we can summarize what has been found in the following way
\begin{itemize}
\item if $\lambda>\delta+\frac{\epsilon}{\delta}\gamma_{F}$, the trajectory leaves ${\mathcal R}_{FREE}$ for $I<I_2$ and then it comes back into ${\mathcal R}_{FREE}$ for $I>I_1$ (with $I_1>I_2$). This implies the possible existence of trajectories turning in counterclockwise direction without ever falling into the sliding zone $(I_2,I_1)$. From previous results, it easily holds that $I_{MAX}>I_1>I_0>I_2>I_{FR}>0$ where $I_2>I_{FR}$; 
%can be easily verified by direct computations
\item if $\delta+\frac{\epsilon}{\delta}\gamma_{F}\ge\lambda\ge\delta+\frac{\epsilon}{\delta}\gamma_{L}$ the sliding zone is reached after at most one sliding line crossing;
\item if $\delta+\frac{\epsilon}{\delta}\gamma_{L}>\lambda>\lambda_4$, the trajectory leaves ${\mathcal R}_{FREE}$ for $I>I_2$ and then it comes back into ${\mathcal R}_{FREE}$ for $I<I_1$ (with $I_2>I_1$). This implies the possible existence of trajectories turning in clockwise direction without ever falling into the sliding zone $(I_1,I_2)$ (this case includes situations in which the sliding line is either a segment or an half straight line). Previous results show that %It easily holds (by looking at the previous sections) 
$I_{MAX}>I_2>I_0>I_{FR}>I_1>0$;
\item if $0<\lambda<\lambda_4$ the sliding zone is reached after at most one sliding line crossing.
\end{itemize}

\begin{figure}[h]
	\begin{center}
		\begin{tabular}{c}
			\hspace{-.2in}
			{ \includegraphics[scale=0.45]{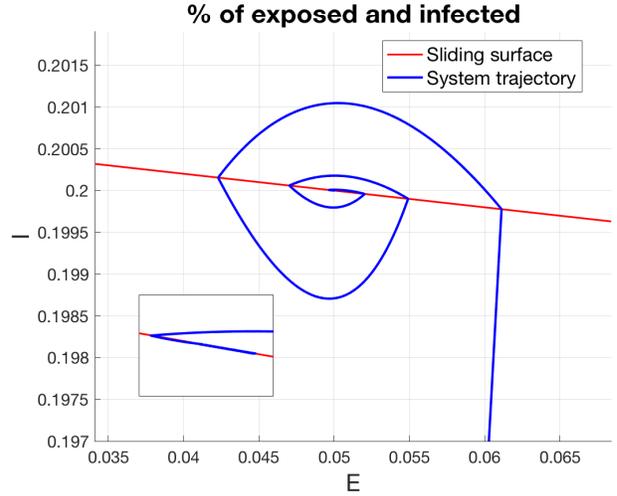}} 
		\end{tabular}
		\caption{{\bf Multiple crossings example} State trajectories on the $(E,I)-$plane with $\lambda=10,\delta=0.05,\epsilon=0.2, \gamma_F=0.065$ and $I_0=2\times10^{-3}$. The condition $\lambda>\delta+\frac{\epsilon}{\delta}\gamma_F$ for having 
multiple crossings of the sliding line is satisfied and its effects are visible in the figure. Crossings appear before the trajectory eventually enters the sliding line. Then, it remains over it forever and moves towards the equilibrium point, as also clearly visible in the zoom on the left bottom.} \label{Fig2}
	\end{center}
\end{figure}

\noindent If $\delta+\frac{\epsilon}{\delta}\gamma_{L}>\lambda>\lambda_4$ the sliding zone is reached after crossing at most two times the sliding line, in view of the position of $I_{FR}$ (inside the sliding zone). In fact, assume (as worst case) that the trajectory starts in ${\mathcal R}_{FREE}$ and crosses the sliding line at some $I_3>I_2$, then coming back to ${\mathcal R}_{FREE}$ at some $I_4<I_1$ (another possibility would be that the trajectory directly falls into the sliding zone). Then, the trajectory cannot touch the line $aw_{FR}, \ a \ge 0$, otherwise this would contradict the unicity of the differential equations solution. In fact, by contradiction, let $P$ be a point in which the latter line is touched. Then two solutions, starting one from the point corresponding to $I_4$ and the other from any point of the form $aP$ with $0<a<1$, would reach the same point $P$, and this is impossible (the solution is unique also by reverting the time direction). In other words, two different solutions cannot intersect, unless they are the same (periodic) one. So the trajectory coming from $I_4$ necessarily would fall on the sliding line into a point included in the interval $(I_1,I_{FR})$, which is a subset of the sliding zone $(I_1,I_2)$, and no other crossings at some $I>I_2$ would be possible. This shows that the only case that needs to be further investigated is the first one, i.e. $\lambda>\delta+\frac{\epsilon}{\delta}\gamma_{F}$. In this case the position of $I_{FR}$ does not prevent multiple crossings of the sliding line, see also Fig. \ref{Fig2}.

\subsubsection{Limit cycles}

Now we show that in the latter case two only possibilities are available
\begin{itemize}
\item the sliding mode takes place after some sliding line crossings;
\item a limit cycle corresponding to two points $I_5,I_6$ exists, with $I_2>I_5>I_{FR}$ and $I_{FR}^L>I_6>I_1$, where $I_{FR}^L$ is the $I-$value corresponding to the intersection between the Frobenius eigenvector of $F_{LOCK}$ and the sliding line.
\end{itemize}
Note that the latter intersection does exist because the sliding line is a segment in the considered case. 
Let's assume $I_{FR}^L>I_1$ (the worst case that surely happens for $\lambda$ large enough), otherwise the convergence to the sliding zone after at most two sliding line crossings would be immediate.
%
%{\color{red}One cannot be sure that $I_{FR}^L>I_1$, but this is not important since we are interested in proving the non existence of the limit cycle in the worst case in which $I_{FR}^L>I_1$. Otherwise, by following a reasoning similar to the previous one in case of $\delta+\frac{\epsilon}{\delta}\gamma_{L}>\lambda>\lambda_4$, the convergence to the sliding zone after at most two sliding line crossings would be immediate. Anyway, it can be easily seen that, at least for $\lambda$ large enough, the condition $I_{FR}^L>I_1$ is really satisfied.}

\noindent Now, consider $x(0) \in {\mathcal L}$ with $I(0):=I_l^{(1)}=0<I_{FR}$. Unicity of the differential equations solution ensures that when the trajectory crosses the sliding line under $I_2$ for the second time, the $I-$value is $I_l^{(2)}>I_{FR}$. At the same time, denoting by $I_u^{(1)}<I_{FR}^L$ the first intersection over $I_1$ and by $I_u^{(2)}<I_{FR}^L$ the second one, it holds that $I_u^{(2)}<I_u^{(1)}$, otherwise the trajectory in ${\mathcal R}_{LOCK}$ would lead to a non-admissible intersection with itself. By performing an inductive reasoning, we can build two sequences such that
$$
I_l^{(1)}<I_l^{(2)}<\dots<I_l^{(n)}<\dots, \ I_u^{(1)}>I_u^{(2)}>\dots>I_u^{(n)}>\dots.
$$
If some $I_l^{(n)}$ or $I_u^{(n)}$ entered the sliding zone $(I_2,I_1)$, the sliding mode would be established. Moreover, in this case, it would be easy to verify that any other trajectory, starting from any point in the sliding line, would do the same i view of the absence of intersections between different solutions of the differential equations. So, the sliding establishment would be global. However, another possibility is available: if %the sequences satisfy
$$
I_l^{(1)}<I_l^{(2)}<\dots<I_l^{(n)}<\dots<I_2, \ I_u^{(1)}>I_u^{(2)}>\dots>I_u^{(n)}>\dots>I_1
$$
the sequences would be both monotone and upper/lower bounded, so two limit points
$$
I_l:=\lim_{n\rightarrow+\infty} \ I_l^{(n)}, \ I_u:=\lim_{n\rightarrow+\infty} \ I_u^{(n)}
$$
would exist. Then, it would be easy to realize that a periodic trajectory passing through $I_l$ and $I_u$ would exist. By resorting to a Lyapunov-like reasoning, in what follows we show that no periodic trajectories can exist if $\lambda>\delta+\frac{\epsilon}{\delta}\gamma_{F}$. So, also in this case the sliding establishment is unavoidable, hence proving the desired global asymptotic stability of $x_{eq}$.

\subsubsection{Lyapunov functions in case of a sliding segment}

We develop a theory assuming $\lambda>\delta$, when the sliding line becomes a segment. This will be sufficient for our purposes, as $\lambda>\delta+\frac{\epsilon}{\delta}\gamma_F$ represents a particular case. 
Let's define two quadratic Lyapunov functions as follows
$$
V_i(E,I)=\epsilon(E-E_0)^2+\delta(I-a_iI_0)^2, 
$$
$$
i=F,L \ \mbox{(where $F$ means freedom and $L$ lockdown)}
$$
with $a_i$ to indicate real numbers to be determined. Let's evaluate the time-derivative of both these functions in the corresponding domains, i.e. $\dot{V}_F$ over ${\mathcal R}_{FREE} \cup {\mathcal L}$ and $\dot{V}_L$ over ${\mathcal R}_{LOCK} \cup {\mathcal L}$. One has
$$ 
\dot{V}_i(E,I)=-2(\epsilon E - \delta I)^2+2\epsilon(\gamma_i-\delta)EI
$$
$$
+2\epsilon\delta(1-a_i)I_0E+2\delta(a_i\delta-\gamma_i)I_0I.
$$
In order to analyze the signs of the two derivatives on ${\mathcal L}$, let's parametrize the sliding segment as follows
$$
m:=\frac{\lambda-\delta}{\epsilon}>0 \ \Rightarrow \ E=E_0-mx, \ I=I_0+x, \ x \in \begin{bmatrix}-1 & \frac{\delta}{\lambda-\delta}\end{bmatrix}I_0
$$
from which
$$
\dot{V}_i(E_0-cx,I_0+x)=-2x^2[m^2\epsilon^2+\epsilon(\delta+\gamma_i)m+\delta^2]
$$
$$
+2I_0[a_i\delta(\delta+\epsilon m)-(\delta^2+m\epsilon\gamma_i)]x.
$$
Hence, the choice
$$
a_i=\frac{\delta^2+m\epsilon\gamma}{\delta(\delta+m\epsilon)}=\frac{\delta^2+\gamma_i(\lambda-\delta)}{\lambda\delta}, \ i=F,L
$$
makes $\dot{V}_i<0$ on the whole sliding line (except for the equilibrium point corresponding to $x=0$). Note that $a_i>1$ if and only if $(\lambda-\delta)(\gamma_i-\delta)>0$ which is verified if and only if $\gamma_i>\delta$. This means that $a_F>1$, while $1>a_L>0$. Now, $\dot{V}_i=0$ holds both at $(0,0)$ and at $(E_0,I_0)$, while it is strictly negative when evaluated at the other points of ${\mathcal L}$. 
To simplify notation, ${\mathcal R}_F$ and ${\mathcal R}_L$ will indicate  ${\mathcal R}_{FREE}$ and ${\mathcal R}_{LOCK}$, respectively. Now, we want to show that $V_i<0$ on the whole region ${\mathcal R}_i$ (except for the origin). This requires to investigate the sign of $\dot{V}_i$ on the boundary of ${\mathcal R}_i$ and the possible existence of internal local maximum points, together with the sign at infinity (for $\dot{V}_L$ only).

\noindent By computing the derivatives of $\dot{V}_i$ w.r.t. $E,I$, and setting them to zero in order to find the critical points, we obtain
$$
\begin{bmatrix}2\epsilon^2 & -\epsilon(\delta+\gamma_i) \cr -\epsilon(\delta+\gamma_i) & 2\delta^2\end{bmatrix}\begin{bmatrix}E \cr I\end{bmatrix}=\delta I_0\begin{bmatrix}\epsilon(1-a_i) \cr a_i\delta-\gamma_i\end{bmatrix}.
$$
This leads to the only solution
$$
E=\frac{\delta^2(2\lambda-\delta+\gamma_i)}{(3\delta+\gamma_i)\lambda\epsilon}I_0, \ I=\frac{\lambda\delta+\delta^2+\gamma_i\lambda-\delta\gamma_i}{(3\delta+\gamma_i)\lambda}I_0
$$
which belongs to ${\mathcal R}_F$ if and only if $(\lambda-\delta)(\lambda+\gamma_i)+\delta^2>0$, a condition verified thanks to the initial assumption $\lambda>\delta$. The Hessian matrix is %given (everywhere) by
$$
H=-2\begin{bmatrix}\epsilon^2 & \epsilon\gamma_i \cr \epsilon\gamma_i & \delta^2\end{bmatrix} \ \Rightarrow \ \det H = 4\epsilon^2(\delta+\gamma_i)(\delta-\gamma_i)<0 \ \mbox{(if $i=F$)}.
$$
Therefore, the only critical point is always in ${\mathcal R}_F$, so $\dot{V}_F$ admits a critical point in ${\mathcal R}_F$ which is a saddle point, while $\dot{V}_L$ doesn't admit critical points in ${\mathcal R}_L$. This implies that only a sign analysis on the boundary is required. Since the analysis on the sliding line has been already performed, we need only to investigate the case $E=0$ or $I=0$. It holds that
$$
\begin{array}{lcl}
E=0&\Rightarrow&\dot{V}_i=2\delta I [(a_i\delta-\gamma_i)I_0-\delta I]<0 \\ 
&& \Rightarrow \ I>\left(a_i-\frac{\gamma_i}{\delta}\right)I_0=\frac{\delta-\gamma_i}{\lambda}I_0 \cr
I=0&\Rightarrow&\dot{V}_i=2\epsilon^2 E [(1-a_i)E_0-E]<0 \\ 
&& \Rightarrow \ E>(1-a_i)E_0=\frac{(\lambda-\delta)(\delta-\gamma_i)}{\lambda\epsilon}I_0.
\end{array}
$$
If $i=F$, $\gamma_F>\delta$ and $a_F>1$ imply that for any $I > 0$ and $E > 0$, respectively, they are both satisfied. Therefore $\dot{V}_F$ is negative over the whole boundary of the bounded set ${\mathcal R}_F \cup {\mathcal L}$ (except for $x=0$ and $x=x_{eq}$, where it vanishes), and does not admit local maxima. So, it is negative everywhere, except for two points. If $i=L$, one has $I_{MAX}>\frac{\delta-\gamma_L}{\lambda}I_0$ and $E_{MAX}=\frac{\lambda}{\epsilon}I_0>\frac{(\lambda-\delta)(\delta-\gamma_L)}{\lambda\epsilon}I_0$, as a consequence of $\lambda^2=\lambda\cdot\lambda>\delta(\lambda-\delta)>(\delta-\gamma_L)(\lambda-\delta)$. So, again they are both satisfied for $I \ge I_{MAX}$ and for $E\ge E_{MAX}$, respectively. However, since ${\mathcal R}_L$ is unbounded, one needs to investigate what happens in a neighborhood of $\infty$. By rewriting
$$
\dot{V}_L=-\begin{bmatrix}E & I\end{bmatrix}\begin{bmatrix}2\epsilon^2 & -\epsilon(\delta+\gamma_L) \cr -\epsilon(\delta+\gamma_L) & 2\delta^2\end{bmatrix}\begin{bmatrix}E \cr I\end{bmatrix}+aE+bI
$$
$$
:=-x^TPx+\begin{bmatrix}a & b\end{bmatrix}x
$$
with $a,b$ suitable real numbers, since $P=P^T>0$ it easily follows that $\dot{V}_L<0$ for $\|x\|$ large enough. So $\dot{V}_L$ is negative everywhere in ${\mathcal R}_L \cup {\mathcal L}$ (except for $x=x_{eq}$), by being negative on the boundary, at infinity, and devoid of internal critical points. This proves the existence of $a_F, a_L$ (uniquely determined) which make $\dot{V}_i$ negative definite in the corresponding regions ${\mathcal R}_i \cup {\mathcal L}$.

\subsubsection{Global asymptotic stability for  $\lambda>\delta+\frac{\epsilon}{\delta}\gamma_F$}

Let's rewrite both the sliding line and the Lyapunov functions in terms of $x:=E-E_0, y:=I-I_0$:
$$
V_F(x,y)=\epsilon x^2 +\delta (y-b_F)^2, \ V_L(x,y)=\epsilon x^2 +\delta (y+b_L)^2, \ y=-mx
$$
where $b_F:=I_0(a_F-1)>0, b_L:=I_0(1-a_L)>0, m>0$. We want to analyze the intersections between the level curves of the Lyapunov functions and the sliding line. For $V_F$ it holds that
$$
V_F(x,y)=c^2, \ y=-mx \ \Rightarrow \ y=\frac{b_Fm^2\delta}{\epsilon+\delta m^2} \pm \bar{y}( c )
$$
$$
=y_{CF} \pm \bar{y}( c ) \ \mbox{(if $c^2 \ge \frac{\epsilon\delta b_F^2}{\epsilon+\delta m^2}$)}
$$
where the inequality is concerned with the sign of discriminant of the second order equation in $y$. Here, there exists $c_{MIN}:=\sqrt{\frac{\epsilon\delta b_F^2}{\epsilon+\delta m^2}}>0$ such that real solutions are available only if $c \ge c_{MIN}>0$, $y_{CF}>0$ (the central point of the intersections is positive), and $\bar{y}( c )$ is monotone increasing (and diverging) as a function of $c$. Through a similar reasoning on $V_L$, one obtains that this time the solution is %will be of the form
$$
y=-y_{CL}\pm \bar{y}(d), \ d \ge d_{MIN}>0, \ y_{CL}>0
$$
where $d$ plays for $V_L$ the same role that $c$ plays for $V_F$. Now, assume by contradiction that a limit cycle exists. By resorting to the same terminology adopted in a previous section, let's define the points of the limit cycle as $I_l, I_u$, with $I_l>I_0>I_u$, and accordingly $y_l:=I_l-I_0, \ y_u:=I_u-I_0$. Recalling that the trajectory from $y_l$ to $y_u$ belongs to ${\mathcal R}_L$, that $V_L$ is there decreasing (because of $\dot{V}_L<0$), so that the value of $d$ decreases while passing from the first point to the second one, and that $y_u>0>y_l$, it easily holds that
$$
y_l=-y_{CL}-\bar{y}(d_1), \ y_u=-y_{CL}+\bar{y}(d_2),  
$$
$$
0<\bar{y}(d_2)<\bar{y}(d_1), \ y_u>0>y_l.
$$
This implies
$$
|y_l|=-y_l=y_{CL}+\bar{y}(d_1)>y_{CL}+\bar{y}(d_2)$$
$$
>-y_{CL}+\bar{y}(d_2)=y_u=|y_u|
$$
so that $|y_u|<|y_l|$. By performing an analogous reasoning passing from $y_u$ to $y_l$, with the trajectory now lying on ${\mathcal R}_F$, by very similar arguments we obtain $|y_l|<|y_u|$, so that $|y_l|<|y_u|<|y_l|$. The contradiction $|y_l|<|y_l|$ shows that no periodic trajectories can take place, so proving that the sliding mode is always reached in finite time. Global asymptotic stability then follows.

\end{document}